\documentclass[twocolumn,showpacs,preprintnumbers,amsmath,amssymb]{revtex4}
\usepackage{graphicx}
\usepackage{dcolumn}
\usepackage{bm}

\begin{document}
\title{IBM-1 calculations towards the neutron-rich nucleus $^{106}$Zr}

\author{Stefan Lalkovski$^{1,2,3}$ and P.~Van~Isacker$^3$}

\affiliation{$^1$Faculty of Physics, University of Sofia, Sofia, BG-1164 Bulgaria\\
$^2$School of Environment and Technology, University of Brighton, Brighton BN2 4JG, UK\\
$^3$Grand Acc\'el\'erateur National d'Ions Lourds,
CEA/DSM--CNRS/IN2P3, BP~55027, F-14076 Caen Cedex 5, France}
\date{\today}

\begin{abstract}
The neutron-rich $N=66$ isotonic and $A=106$ isobaric chains,
covering regions with varying types of collectivity,
are interpreted in the framework of the interacting boson model.
Level energies and electric quadrupole transition probabilities
are compared with available experimental information.
The calculations for the known nuclei in the two chains
are extrapolated towards the neutron-rich nucleus $^{106}$Zr.
\end{abstract}

\pacs{}
\maketitle
\section{Introduction}
\label{s_int}
In the last decade the neutron-rich nuclei
in the $40\leq Z \leq50$ region have attracted
both theoretical and experimental attention.
They were extensively studied
via spontaneous or induced fission reactions.
Nuclei from this region of Segr\'e chart
exhibit vibrational, transitional, and rotational types of collectivity.
The neutron-rich palladium and ruthenium nuclei, for example,
show a typical transitional behavior
while molybdenum isotopes exhibit
a vibrational to near-rotational evolution~\cite{Urban04,Lalkovski05}.
Such changes in the degree of collectivity
are even stronger in the zirconium isotopic chain~\cite{Urban04,Lalkovski05}
where the structure evolves from near-vibrational in $^{98}$Zr
to rotational-like in $^{104}$Zr,
the latter being the most deformed nucleus
of the neutron-rich $Z=40$ isotopes.
As the next zirconium isotope $^{106}$Zr
is an $N=66$ mid-shell nucleus,
an even larger deformation can be expected. 

A detailed analysis of some spectroscopic observables,
such as the ratio $R_{4/2}$ of excitation energies
of the first $2^+$ and $4^+$ levels~\cite{Urban04,Lalkovski05}
or the amplitude of even-odd staggering in the $\gamma$ band~\cite{Lalkovski05},
shows that mid-shell effects in palladium isotopes
occur in $^{114,116}$Pd,
{\it i.e.} two and four neutrons beyond $N=66$.
In the ruthenium isotopes these effects arise precisely at $N=66$
while in the molybdenum chain
they occur in $^{106}$Mo, {\it i.e.} two neutrons before mid-shell. 
Furthermore, it was shown that the degree of collectivity decreases
in the heavier $N=66,68$ molybdenum isotopes $^{108,110}$Mo.
In spite of the fact that the experimental information
on the heavy zirconium nuclei is poor,
it is clear that the degree of collectivity increases
towards $^{104}$Zr~\cite{Urban04} 
which has 64 neutrons
and is the heaviest zirconium isotope known to date.
The shift of the mid-shell in the various isotopic chains
was explained in Ref.~\cite{Urban04}
as resulting from the filling of the neutron $h_{11/2}$ and proton $g_{9/2}$ intruder orbitals. 

Self-consistent calculations show
that in the region of neutron-rich nuclei approaching the neutron-drip line,
the single-particle shell structure may significantly change 
due to the diffuseness of the neutron density~\cite{Dobaczewski94}.
Also, as widely discussed in the literature,
due to a predicted weakening of the spin-orbit force, 
new magic numbers may be expected in neutron-rich nuclei.
For instance, in a study of $N=82$ isotones,
evidence for shell quenching was presented in Ref.~\cite{Kaurzsch00}.
On the other hand, Jungclaus {\it et al.}~\cite{Jungclaus07} argued
that from the isomeric decay study in $^{130}$Cd
there is no direct evidence for the shell quenching
in the region of the heavy cadmium nuclei.

The lower-$Z$ $N=82$ nuclei, however, have not yet been studied experimentally
and the suppression of the shell effects in this region is still an open question.
Properties of these nuclei were predicted 
within a relativistic Hartree-Fock Bogoliubov (HFB) approach~\cite{Sharma02}.
The calculations, performed with different parametrizations,
show that the $N=82$ shell gap
persists in the heavy palladium and ruthenium nuclei
but that a weakening of this gap is expected in the zirconium isotopes.
For example, the two-neutron separation energies from $^{118}$Zr to $^{126}$Zr,
obtained via HFB calculations with the SkP interaction,
show no discontinuity at $N=82$ but a rather smooth behavior.
As discussed above,
nuclear collective behavior is expected to enhance towards the mid-shell
as a result of the increase in valence particle number.
If, however, the HFB+SkP scenario for the zirconium isotopes turns out to be valid
and $N=82$ vanishes or is weakened in the neutron-rich region,
then the determination of valence-particle number is fraught with ambiguity,
leading to a different behavior of nuclear collectivity in this isotopic chain.
The aim of the present paper, therefore,
is to predict the spectroscopic properties
concerning the $N=66$ mid-shell zirconium nucleus $^{106}$Zr, 
based on the assumption that $N=82$
remains a magic number in the heavy zirconium isotopes.
Comparison with results of future experiments on this nucleus
will then reveal whether this hypothesis is borne out or not.

\section{Model}
\label{s_mod}
Often the only available information about neutron-rich nuclei 
comes from prompt $\gamma$-ray spectroscopy.
The observed levels are grouped into $\Delta J=2$ sequences,
corresponding to the ground-state band,
and/or into $\Delta J=1$ sequences, based on a $2^+_2$ level
which usually is interpreted as a quasi-$\gamma$ band~\cite{Sakai84}. 
Such structures naturally appear in the framework
of the interacting boson model (IBM)~\cite{Iachello87}
which has been shown to be successful
in the description of nuclear collective properties.

The IBM in its first version, known as \mbox{IBM-1},
is based on the assumption that nuclear collectivity
can be expressed in terms of $s$ and $d$ bosons~\cite{Iachello87}.
The model Hamiltonian is constructed from a set of 36 operators,
bilinear in the boson creation and annihilation operators
and generating the U(6) Lie algebra.
Dynamical symmetries occur if the Hamiltonian
can be written as a combination
of invariant (or Casimir) operators of specific subalgebras of U(6)~\cite{Iachello06}
and three such cases occur,
namely the spherical vibrational limit U(5),
the deformed limit SU(3),
and $\gamma$-soft limit SO(6).
These dynamical symmetries generate energy spectra with states
that are labeled by the irreducible representations
of the algebras in the respective chains
that reduce the dynamical algebra U(6)
into the symmetry algebra SO(3) of rotations in three-dimensional space. 
The different limits thus correspond to nuclei
with distinct collective properties.
For example, in the SO(6) limit
the ground-state and $\gamma$ bands
lie in the same representation
while in the SU(3) limit they are in different ones,
leading to forbidden or weaker inter-band transitions in the latter case. 

In the $A\approx110$ nuclei of interest here
strong transitions between the two bands are observed
indicating that none of these nuclei can be interpreted
in the exact SU(3) limit,
but rather a transitional behavior should be expected.
This can be achieved in the \mbox{IBM-1}
by the use of the full Hamiltonian
which reads~\cite{Iachello87}
\begin{equation}
\hat H=\epsilon_d\hat n_d+
\kappa\,\hat Q^\chi\cdot\hat Q^\chi+
\kappa'\hat L\cdot\hat L+
c_3\hat T_3\cdot\hat T_3+
c_4\hat T_4\cdot\hat T_4,
\label{e_ham}
\end{equation}
where $\hat n_d\equiv d^\dag\cdot\tilde d$
and $\hat L_\mu\equiv\sqrt{10}[d^\dag\times\tilde d]^{(1)}_\mu$
are the $d$-boson number
and the angular momentum operators, respectively.
Furthermore, the quadrupole operator is defined as
$\hat Q^\chi_\mu\equiv[d^\dag\times\tilde s]^{(2)}_\mu+\chi[d^\dag\times\tilde d]^{(2)}_\mu$,
while the last two terms in the Hamiltonian involve the operators
$\hat{T}_{3,\mu}\equiv[d^\dagger\times\tilde{d}]^{(3)}_\mu$ and
$\hat{T}_{4,\mu}\equiv[d^\dagger\times\tilde{d}]^{(4)}_\mu$.
In the present work there is no need for the rotational term $\hat L\cdot\hat L$
and we take $\kappa'=0$ throughout.
The total number of bosons $N=n_s+n_d$
is taken as half the number of valence particles or holes,
counted from the nearest closed-shell configuration,
following the prescription of Ref.~\cite{Casten90}.
As such, \mbox{IBM-1} calculations
are indirectly related to the underlying shell structure.
Finally, in the consistent-$Q$ formalism~\cite{Warner82}
the operator for electric quadrupole transitions
is introduced as $\hat T_\mu({\rm E2})=e_{\rm b}\hat Q^\chi_\mu$,
where $e_{\rm b}$ is the boson effective charge
and $Q^\chi_\mu$ is the quadrupole operator
with the same parameter $\chi$
as in the Hamiltonian~(\ref{e_ham}).  

\section{Results and discussion}
\label{s_resdis}
Since the addition or subtraction of a few nucleons 
may change significantly a nuclear spectrum,
two nuclei with different numbers of valence neutrons and protons
but with the same {\em total} number of valence nucleons---and
hence the same number of bosons---may
display different types of collectivity.
In adjusting the parameters of the Hamiltonian~(\ref{e_ham})
to spectra observed in a given region of the nuclear chart,
it is therefore not sufficient to assume them to be constant for all nuclei
since that would lead to identical spectra for nuclei with the same $N$.
Instead, some dependence of the parameters
on the {\em separate} neutron and proton boson numbers $N_\nu$ and $N_\pi$
should be imposed.
To avoid the complexity of determining
the correct functional dependence on $N_\nu$ and $N_\pi$
for each of the Hamiltonian parameters,
we have followed the simpler procedure
of separately considering chains of isotopes, isotones, or isobars
that end with the nucleus of interest, $^{106}$Zr,
and fitting each chain with an independent set of parameters.
This method avoids the problem of having identical spectra
for constant boson number $N$ (such nuclei belong to different chains)
and, in addition, since the extrapolation to $^{106}$Zr
obtained from the different chains may vary,
it gives an idea on the possible error of the prediction.

\begin{figure*}
\begin{center}
\includegraphics[width=12cm]{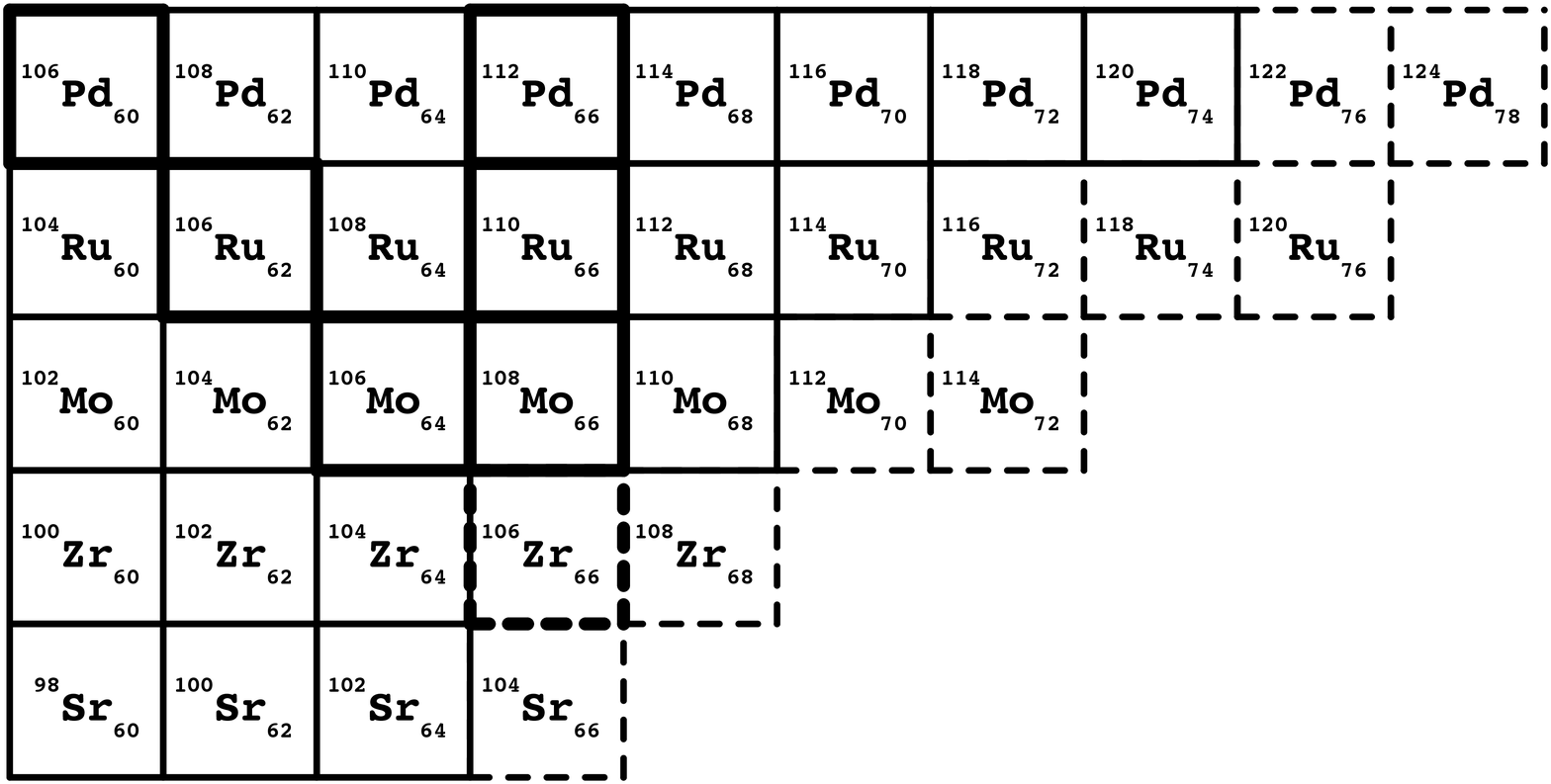}
\caption{
Even-even nuclei in the region close to $^{106}$Zr.
All nuclei in boxes with full lines have known excited levels
while those in boxes with broken lines
are only known in their ground state.
The nuclei used in the fit are in boxes with thick lines.}
\label{f_chart}
\end{center}
\end{figure*}
In the present application to $^{106}$Zr,
the $N=66$ isotonic chain includes $\gamma$-unstable $^{112}$Pd,
more triaxial $^{110}$Ru, and near-rotational $^{108}$Mo.
The second set of nuclei belongs to the $A=106$ isobaric chain~\cite{Frenne08}
extending from vibrational-like $^{106}$Pd to near-rotational $^{106}$Mo.
Unfortunately, the zirconium isotopic chain does not allow
the determination of a unique set of model parameters
since only yrast data are available
for the neutron-rich isotopes $^{100,102,104}$Zr.
Consequently, only the isotonic and isobaric chains are considered in the fit.
Nuclei in a given chain are distinguished
by the number of bosons ({\it i.e.}, particle or hole pairs),
counted from the nearest closed-shell configurations
which are $Z=28$ or $Z=50$ for the protons,
and $N=50$ or $N=82$ for the neutrons.
Figure~\ref{f_chart} summarizes known nuclei in the neighbourhood of $^{106}$Zr.
Nuclei with known excited levels are distinguished (full lines)
from those of which only ground-state properties are known (dashed lines).
Furthermore, the nuclei used in the fit are high-lighted with thick lines.

The model parameters $\epsilon_d$, $\kappa$, $c_3$, and $c_4$,
listed in Table~\ref{t_param},
are obtained after a least-squares fit to the experimental level energies.
\begin{table}
\caption{
Parameters and rms deviation (in keV)
for the $N=66$ isotones and the $A=106$ isobars.}
\label{t_param}
\begin{ruledtabular}
\begin{tabular}{crrrrrr}
&$\epsilon_d$&$\kappa$&$c_3$&$c_4$&$\chi\footnotemark[1]$&rms\\
\hline
$N=66$&   1171& $-24.3$& $-70.8$& $-134.9$& $-0.30$& 58\\
$A=106$& 1053& $-25.6$& $-70.6$& $-118.3$& $-0.30$& 143
\end{tabular}
\footnotetext[1]{Dimensionless.}
\end{ruledtabular}
\end{table}
Since its effect on the excitation energies is weak,
the parameter $\chi$ is varied independently
to give a fair description of E2 transition rates in $^{106}$Pd
within the consistent-$Q$ formalism,
and other nuclei are calculated with the same value.
Although within a given chain parameters may vary with boson number,
any allowed variation does not markedly improve the quality of the fit.
Consequently, to keep the number of parameters to a minimum,
they are taken constant for a given chain
and any structural evolution within that chain
is due to the changing boson number $N$.

The quality of the fit is summarized
by the root-mean-square (rms) deviation
between the experimental and calculated level energies.
The rms values obtained in the two independent fits
are given in Table~\ref{t_param}.
The experimental~\cite{Frenne08,Krucken01,Frenne00,Zhuo03,Blachot00}
and theoretical energies are listed
in Tables~\ref{t_exn66} and~\ref{t_exa106} for the $N=66$ and $A=106$ chains,
respectively.
The nucleus $^{106}$Pd is known in greater detail
and is shown separately in Table~\ref{t_expd106}.
\begin{table}
\caption{Experimental and theoretical excitation energies (in keV)
of levels in neutron-rich $N=66$ isotones.}
\label{t_exn66}
\begin{ruledtabular}
\begin{tabular}{crrrrrrrr} 	
nucleus&
$J^\pi_{\rm gsb}$& $E_{\rm gsb}^{\rm ex}$& $E_{\rm gsb}^{\rm th}$&
$J^\pi_\gamma$ & $E_\gamma^{\rm ex}$& $E_\gamma^{\rm th}$\\
\hline
$^{112}$Pd &$0^+$  &       0&       0&   $2^+$&   736&   775\\
($N=10$)   & $2^+$  &   348&   354&   $3^+$& 1096& 1138\\
                  & $4^+$  &   883&   849&   $4^+$& 1362& 1304\\
                  & $6^+$  & 1550& 1481&   $5^+$& 1759& 1692\\
                  & $8^+$  & 2318& 2240&   $6^+$& 2002& 1934\\
                  & $10^+$& 3049& 3118&   $7^+$& 2483& 2357\\
                  &             &         &         &   $8^+$& 2638& 2669\\
                  &             &         &         &   $9^+$& 3085& 3129\\
                  &             &         &         & $10^+$& 3327& 3507\\
\hline
 $^{110}$Ru&  $0^+$&       0&       0&   $2^+$&   613&   689\\
($N=11$)    &   $2^+$&   241&   253&   $3^+$&   860&   960\\
                   &   $4^+$&   663&   671&   $4^+$& 1084& 1139\\
                   &   $6^+$& 1239& 1239&   $5^+$& 1375& 1449\\
                   &   $8^+$& 1945& 1945&   $6^+$& 1684& 1699\\
                   & $10^+$& 2759& 2777&   $7^+$& 2021& 2060\\
                   &             &         &         &   $8^+$& 2397& 2377\\
                   &             &         &         &   $9^+$& 2777& 2789\\
                   &             &         &         & $10^+$& 3255& 3168\\
\hline
$^{108}$Mo &   $0^+$&       0&       0&   $2^+$&   586&   647\\
($N=12$)     &   $2^+$&   193&   178&   $3^+$&   783&   830\\
                    &   $4^+$&   564&   533&   $4^+$&   978& 1010\\
                    &   $6^+$& 1090& 1047&   $5^+$& 1232& 1257\\
                    &   $8^+$& 1753& 1705&   $6^+$& 1508& 1505\\
                    & $10^+$& 2529& 2495&   $7^+$& 1817& 1819\\
                    &             &         &         &   $8^+$& 2170& 2132\\
                    &             &         &         &   $9^+$& 2524& 2506\\
                    &             &         &         & $10^+$& 2950& 2880\\
\hline
$^{106}$Zr &   $0^+$&   0&      0&   $2^+$& ---&  618\\
($N=13$)   &   $2^+$& ---&   141&   $3^+$& ---&  754\\
                  &   $4^+$& ---&   455&   $4^+$& ---&  922\\
                  &   $6^+$& ---&   926&   $5^+$& ---&1133\\
                  &   $8^+$& ---& 1543&   $6^+$& ---&1371\\
                  & $10^+$& ---& 2297&   $7^+$& ---&1653\\
                  &             &    &         &   $8^+$& ---&1956\\
                  &             &    &         &   $9^+$& ---&2303\\
                  &             &    &         & $10^+$& ---&2668\\
\end{tabular}
\end{ruledtabular}
\end{table}
\begin{table}
\caption{Experimental and theoretical excitation energies (in keV)
of levels in neutron-rich $A=106$ isobars.}
\label{t_exa106}
\begin{ruledtabular}
\begin{tabular}{crrrrrr} 	
nucleus&
$J^\pi_{\rm gsb}$& $E_{\rm gsb}^{\rm ex}$& $E_{\rm gsb}^{\rm th}$&
$J^\pi_\gamma$&  $E_\gamma^{\rm ex}$&   $E_\gamma^{\rm th}$\\
\hline
$^{106}$Ru&   $0^+$&       0&       0& $2^+$&   792&   772\\
($N=9$)      &   $2^+$&   270&   362& $3^+$& 1092& 1156\\
                   &   $4^+$&   715&   865& $4^+$& 1307& 1308\\
                   &   $6^+$& 1296& 1498& $5^+$& 1641& 1720\\
                   &   $8^+$& 1973& 2252& $6^+$& 1908& 1943\\
                   & $10^+$& 2705& 3115& $7^+$& 2284& 2387\\
                   &             &         &         & $8^+$& 2960& 2676\\
\hline
$^{106}$Mo&   $0^+$&       0&       0&   $2^+$&   711&   651\\
($N=11$)    &   $2^+$&   172&   205&   $3^+$&   885&   881\\
                   &   $4^+$&   522&   582&   $4^+$& 1068& 1057\\
                   &   $6^+$& 1033& 1110&    $5^+$& 1307& 1338\\
                   &   $8^+$& 1688& 1773&   $6^+$& 1563& 1578\\
                   & $10^+$& 2472& 2559&   $7^+$& 1868& 1916\\
                   &             &         &         &   $8^+$& 2194& 2217\\
                   &             &         &         &   $9^+$& 2559& 2608\\
                   &             &         &         & $10^+$& 2951& 2965\\ 
\hline
$^{106}$Zr &   $0^+$&    0&      0&   $2^+$&  ---&  630\\
($N=13$)   &   $2^+$&  ---&   134&   $3^+$&  ---&  762\\
                  &   $4^+$&  ---&   433&   $4^+$&  ---&  924\\
                  &   $6^+$&  ---&   883&   $5^+$&  ---&1129\\
                  &   $8^+$&  ---& 1473&   $6^+$&  ---&1357\\
                  & $10^+$&  ---& 2192&   $7^+$&  ---&1630\\
                  &             &     &         &   $8^+$&  ---&1919\\
                  &             &     &         &   $9^+$&  ---&2254\\
                  &             &     &         & $10^+$&  ---&2601\\
\end{tabular}
\end{ruledtabular}
\end{table}
\begin{table}
\caption{Experimental and theoretical excitation energies (in keV)
of levels in $^{106}$Pd.}
\label{t_expd106}
\begin{ruledtabular}
\begin{tabular}{crrrrrr} 	
nucleus& $J^\pi$& $E_1$& $E_2$& $E_3$& $E_4$& $E_5$\\
\hline
$^{106}$Pd& $0^+_{\rm ex}$&        0& 1134& 1706& 2001& 2278\\
($N=7$)     &  $0^+_{\rm th}$&        0& 1017& 1384& 2080& 2297\\
                  & $2^+_{\rm ex}$&    512& 1128&  1562& 1909& 2242\\
                  &  $2^+_{\rm th}$&    519&   982&  1575& 1859& 2004\\
                  & $3^+_{\rm ex}$&  1558&&&&\\
                  &  $3^+_{\rm th}$&  1483&&&&\\
                  & $4^+_{\rm ex}$&  1229& 1932& 2077& 2283&\\
                  &  $4^+_{\rm th}$&  1153& 1617& 2010& 2213&\\
                  & $5^+_{\rm ex}$&  2366&&&&\\
                  &  $5^+_{\rm th}$&  2141&&&&\\
                  & $6^+_{\rm ex}$&  2077&&&&\\
                  &  $6^+_{\rm th}$&  1896&&&&\\
\end{tabular}
\end{ruledtabular}
\end{table}
In both the isotopic and isobaric chains,
the observed energy of the $2^+_1$ state
decreases when approaching the nucleus $^{106}$Zr.
The same trend is observed for the calculated energies
as a function of boson number.
The experimental energy of the $\gamma$-band head
also decreases with increasing boson number
and a similar evolution is observed for the $2^+_2$ theoretical energies.
According to the systematics of the less neutron-rich nuclei,
the $2^+_1$ level in $^{106}$Zr should thus lie around 140~keV
while the $2_2^+$ level should be in the region $600$--650~keV.
Furthermore, the two different fits
give approximately the same $\gamma$-band energies.
Even though the two fitted parameter sets are close (see Table~\ref{t_param}),
the $N=66$ set gives rise to a slightly faster increase
of the ground-state band energies
than is the case for the $A=106$ set.

\begin{figure*}
\begin{center}
\includegraphics[width=17cm]{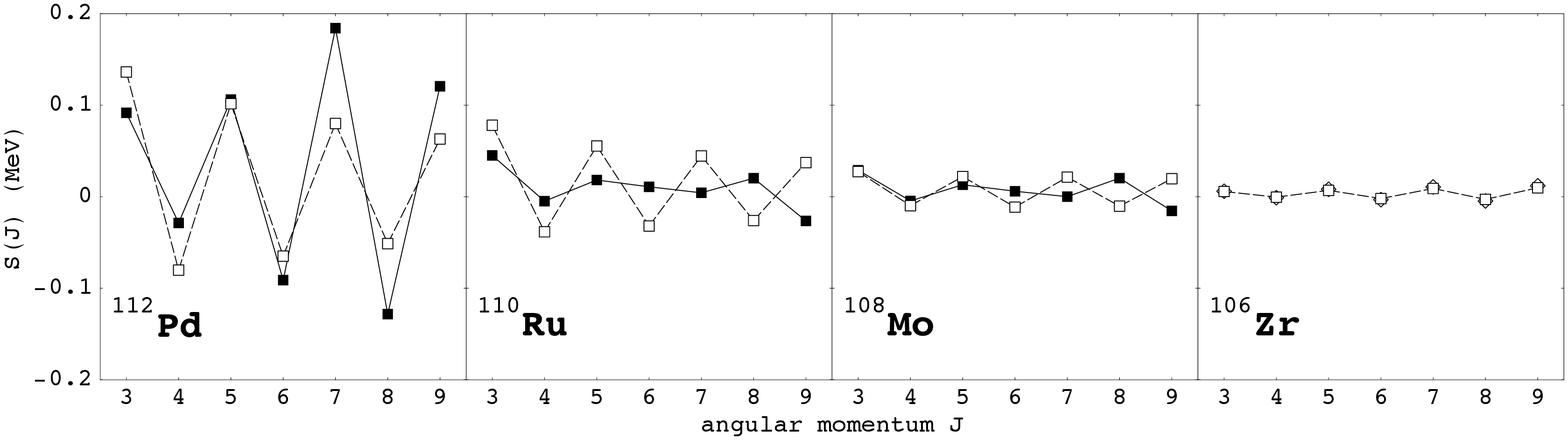}
\caption{
Experimental (full symbols connected with full lines)
and theoretical (open symbols connected with dashed lines)
$\gamma$-band staggering $S(J)$ in the $N=66$ isotones.
In $^{106}$Zr both near-identical predictions are shown
based on the extrapolation of the $N=66$ isotones (open squares)
and of the $A=106$ isobars (open diamonds).}
\label{f_stag}
\end{center}
\end{figure*}
The displacement of odd-spin with respect to even-spin $\gamma$-band levels
can be quantified with the following function:
\begin{equation}
S(J)=E(J)-\frac{(J+1)E(J-1)+JE(J+1)}{2J+1},
\end{equation}
where  $E(J)$ is the energy of the level with angular momentum $J$.
The experimental and theoretical staggering in the $N=66$ isotones
is plotted in Fig.~\ref{f_stag};
to gauge any systematic behavior, plots are drawn to the same scale.
In general, the staggering amplitude is high for spherical nuclei
and decreases towards the mid-shell region.
The $N=66$ isotones do not constitute an exception
to the systematic trends established in Ref.~\cite{Lalkovski05}.
The $\gamma$-unstable nucleus $^{112}$Pd (10 bosons)
shows the highest amplitude in $S(J)$
among the $N=66$ isotones.
The $\gamma$-band staggering
is less pronounced in $^{110}$Ru (11 bosons)
while in $^{108}$Mo (12 bosons) it is highly suppressed.
From these systematics trends a $\gamma$ band
with a very low staggering amplitude
is expected in $^{106}$Zr,
namely $S(J)\approx10$~keV,
which would make $^{106}$Zr the nucleus with the lowest staggering amplitude
observed in the $40\leq Z\leq50$ region. 

\begin{table}
\caption{Theoretical $B$(E2) values (in units e$^2$b$^2$) for transitions
in neutron-rich $N=66$ isotones.}
\label{t_e2n66}
\begin{ruledtabular}
\begin{tabular}{cllll} 
transition& $^{112}$Pd& $^{110}$Ru& $^{108}$Mo& $^{106}$Zr\\
\hline
$2_1^+\rightarrow0_1^+$&   0.27&   0.35&   0.45& 0.53\\
$4_1^+\rightarrow2_1^+$&   0.43&   0.54&   0.65& 0.75\\
$0_2^+\rightarrow2_1^+$&   0.22&   0.17& 0.057& 0.003\\
$2_2^+\rightarrow0_1^+$& 0.015& 0.024& 0.025& 0.020\\
\end{tabular}
\end{ruledtabular}
\end{table}
\begin{table}
\caption{Experimental and theoretical $B$(E2) values (in units e$^2$b$^2$) for transitions
in neutron-rich $A=106$ isobars.}
\label{t_e2a106}
\begin{ruledtabular}
\begin{tabular}{clllll} 
transition&\multicolumn{2}{c}{$^{106}$Pd}& $^{106}$Ru& $^{106}$Mo& $^{106}$Zr\\
\cline{2-3}
&Expt&Theo&Theo&Theo&Theo\\
\hline 
$2_1^+\rightarrow0_1^+$&   0.134~(4)&     0.14&     0.23&   0.37& 0.54\\
$4_1^+\rightarrow2_1^+$&   0.211~(17)&   0.22&     0.36&   0.55& 0.77\\
$0_2^+\rightarrow2_1^+$&   0.105~(20)&   0.13&     0.16& 0.092& 0.0044\\
$2_2^+\rightarrow0_1^+$& 0.0035~(3)& 0.0019& 0.0097& 0.023& 0.021\\
\end{tabular}
\end{ruledtabular}
\end{table}
Also electric quadrupole transition probabilities
have been considered for the neutron-rich nuclei
belonging to the $N=66$ and $A=106$ chains.
The calculated values are listed in Tables~\ref{t_e2n66} and~\ref{t_e2a106}
along with the few known experimental values~\cite{Frenne08}. 
The boson effective charge $e_{\rm b}=0.11$~eb
is obtained after a least-squares fit to the $B$(E2) values
measured in $^{106}$Pd
and this value is used for all other nuclei.
The results show an increase of the $B({\rm E2};2_1^+\rightarrow0_1^+)$ value
from the near-vibrational $^{106}$Pd
towards the near-rotational $^{108}$Mo.
This $B({\rm E2})$ value is predicted highest for $^{106}$Zr
and the calculations for the two different chains
give a consistent value of
$B(E2;2_1^+\rightarrow0_1^+)\approx0.50$--0.55~e$^2$b$^2$.
Furthermore, we note from Table~\ref{t_e2a106}
the behavior predicted for the $B(E2;2_2^+\rightarrow0_1^+)$ value
which is small in $^{106}$Pd and $^{106}$Zr
but becomes large for the isotopes in between,
a characteristic feature of transitional nuclei.

\section{Conclusions}
The aim of this paper was to illustrate
the usefulness of the interacting boson model in its simplest version, the \mbox{IBM-1},
for predicting properties of exotic nuclei.
The usual difficulty that arises with such attempts
is that the variations of the model's parameters
with the valence neutron and proton numbers are not known,
precluding a reliable extrapolation to unknown regions of the nuclear chart.
Our method proposes to circumvent this problem
by studying the structural evolution in three different chains of nuclei,
namely isotopic, isotonic, and isobaric ones,
which cross at the nucleus of interest
at the outskirts of the region of stable nuclei.
Predictions are obtained by extrapolating
the different chains to the exotic nucleus in question,
and, in addition, a comparison of these extrapolations
gives an idea of the errors involved.

This method was applied to the neutron-rich members
of the $N=66$ isotonic and $A=106$ isobaric chains
of which level energies and electric quadrupole transition probabilities
were fitted with \mbox{IBM-1}.
The two chains intersect at $^{106}$Zr
which allowed the prediction of this nucleus' excitation energies
and electric quadrupole transition properties.

As a final remark we emphasize
that the \mbox{IBM-1} is a valence-nucleon model
and that the extrapolations as described here 
crucially depend on the definition of neighboring closed-shell configurations.
Thus, the results on $^{106}$Zr indirectly involve
the assumption of the persistence of magic numbers
in this region of neutron-rich nuclei.

\section{Acknowledgments}
S.L. acknowledges the Bulgarian National Science Fund
for a Research Fellowship Grant.
This work was partially supported by the
Agence National de Recherche, France,
under contract nr ANR-07-BLAN-0256-03.

\end{document}